\documentclass[twocolumn,floatfix,pra,showpacs,superscriptaddress]{revtex4-1}
\usepackage{helvet}
\usepackage{eurosym}
\usepackage{color}
\usepackage[latin1]{inputenc}
\usepackage{subfigure}
\usepackage{amssymb}
\usepackage{amsmath}
\usepackage{graphicx}
\usepackage{bbold}
\usepackage{natbib}

\DeclareGraphicsExtensions{%
    .png,%
    .jpg,%
    .pdf,.PDF,%
    .mps,.jpeg,.jbig2,.jb2,.JPG,.JPEG,.JBIG2,.JB2}

\newcommand{\avg}[1]{\left< #1 \right>} 

\begin{document}
\title{Time delayed control of excited state quantum phase transitions in the Lipkin-Meshkov-Glick model}
\author{Wassilij Kopylov}
\affiliation{Institut f\"ur Theoretische Physik,
  Technische Universit\"at Berlin,
  D-10623 Berlin,
  Germany}
\author{Tobias Brandes}
\affiliation{Institut f\"ur Theoretische Physik,
  Technische Universit\"at Berlin,
  D-10623 Berlin,
  Germany}
\date{\today}

\begin{abstract}
We investigate the role of dissipation in excited state quantum phase transitions (ESQPT) within the Lipkin-Meshkov-Glick model. Signatures of the ESQPT are directly visible in the complex spectrum of an effective Hamiltonian, whereas they get smeared out in the time-dependence of system observables. In the latter case, we show how delayed feedback control can be used to restore the visibility of the ESQPT signals.  
\end{abstract}
\maketitle

\section{Introduction}
The properties of a quantum phase transition (QPT) \cite{Sachdev-QPT}  and an excited state QPT (ESQPT) \cite{ESQPT-inmany_body_systems-Caprio,ESQPT-in_systesm_with_2_degrees-Cejnar} driven by quantum fluctuations in many body quantum systems at zero temperature feeds the interest since many decades. At a quantum level, the ESQPT is hidden in the system's density spectrum as discontinuities or divergence \cite{ESQPT-inmany_body_systems-Caprio,brandes2013excited}. For some famous models like Dicke superradiance  \cite{Dicke-Dicke_Modell} or Lipkin-Meshkov-Glick (LMG) \cite{LMG-lipkin1965validity}, the main properties of both QPTs can be explained and understood at a semiclassical level in the thermodynamic limit, where a QPT corresponds to a bifurcation \cite{ESQPT-quantum-quench-pedro,Bhaseen_dynamics_of_nonequilibrium_dicke_models} and a ESQPT is connected to a saddle point in the semi-classical energy potential \cite{ESQPT-inmany_body_systems-Caprio}. Furthermore, the density of states can then be calculated analytically \cite{brandes2013excited,LMG-spectrum_thermodynamic_limit_and_finite_size-corr-Mosseri}.  For both models, the ESQPT manifests in the observable average as a peak at a certain energy \cite{ESQPT_in_quantum_optical_models-Pedro,ESQPT_Docoherence_two_level_boson_pedro,brandes2013excited} and was already observed in molecular systems \cite{ESQPT-exp_molecular_system-Koput} or microwave Dirac billiards \cite{ESQPT-Dirac_billiards-Dietz}. 

Feedback control is a promising tool to change the system dynamics in a desired way. Influence of the laser statistics \cite{yamamoto_statistic_feedback_and_laser}, neurosystems \cite{schoell_time_del_in_neurosystems} or even control at a quantum level \cite{Wiseman-Quantum_measurment_control,Dicke-nonequilibrium_qpt-bastidas,Poeltl_pure_state_stability_by_Feedback,Kabuss-quantum_feedback_anal_study_and_rabi_osci,Grimsmo-time_delayed_quantum_feedback_control} are only some examples of its powerfulness. One usually distinguishes between the closed and open control loops, in the last one the feedback depends on the state of the system, where time delayed Pyragas control \cite{pyragas1992continuous} has an important niche. 
For instance, it was used to speed up the convergence to a steady state in the dissipative quantum system \cite{Dicke_Rapid_convergence_time_delay-Grimsmo}. 

In our previous works, we applied time delayed Pyragas control to a Dicke model \cite{Dicke-Dicke_Modell} to create new non-equilibrium phases \cite{Kopylov-time_delayed_control_Dicke} and suggested a new method to extract the ESQPT-signal from the time evolution in the closed LMG system \cite{LMG-lipkin1965validity,LMG-TC-periodic_dynamic_and_QPT-Georg}. Furthermore, closed loop control was already applied to the LMG model to induce new phases or to modify the divergence in density of states \cite{Bastidas-Critical_quasienergy_in_driven_systems,Bastidas-Quantum_criticality_in_kicked_top,LMG-ac_driven_QPT-Georg}.  Moreover, Oberthaler et al. implemented the LMG model using the existing ingenious experimental setup based on $^{87}Rb$ Bose-Einstein condensates \cite{LMG-Exp_Bifurcation_rabi_to_jesophson-Oberthaler}, which offer a high degree of freedom for system preparation, providing a possibility to test new insights about the (ES)QTP. Surprisingly, the effects of a dissipative environment on the ESQPT seems not to be well studied in contrast to the QPT \cite{Morrison-Dissipative_LMG-and_QPT}, though they are always present in experimental realisations.  
 
Inspired by this, we study the effects of dissipation on the ESQPT signal in the LMG model and apply a time delayed feedback scheme to cancel them by the creation of new phases. In our model, we condition the atomic coupling on the difference of a spin observable average at two different times and perform the calculation at a mean-field-level. On top we show, that in dissipative systems the ESQPT is directly visible from a complex system spectrum.

Our work is organised as follows. In section II we introduce the dissipative LMG model and the feedback. In section III we study the dissipative effects on a ESQPT at a quantum level evaluating the complex spectrum of an effective Hamiltonian. In section IV we show the smoothing effects of the ESQPT signal due to dissipation at a mean-field level and show in section V how to compensate them using time delayed Pyragas control. In the last VI section we discuss the results.


\section{LMG Model with Dissipation and Feedback}
The LMG model \cite{LMG-lipkin1965validity} describes an interaction between $N$ spins and represents a special case of a Heisenberg Model. In general, the LMG Hamiltonian reads 
\begin{equation}\label{eq:LMG-Hamiltonian}
\hat{H} = -h \hat{J}_z - \frac{\gamma_x}{N} \hat{J}_x^2 - \frac{\gamma_y}{N} \hat{J}_y^2 , 
\end{equation}
where $\hat{J}_{i} = \sum_{j=1}^{N} \hat{\sigma}_i^j, i \in \{x,y,z\}$ are collective angular momentum operators, $h$ is an effective parameter for external magnetic field in $z$-direction and $\gamma_{x}$ or $\gamma_{y}$ describes the spin-spin interaction strength, which is the same for all spins in the LMG model. In the following, we will always use the isotropic $\gamma_{y} = 0$ case, if we do not explicitly point to  $\gamma_{y} \neq 0$ case.   \\
Especially in the experimental realizations, the LMG system is always coupled to an environment which cause the damping and thermalisation of the system and can be modelled by a master equation \cite{Morrison-Dissipative_LMG-and_QPT,Morrison-Collective_spin_system-QPT_and_entaglement} with collective decay \cite{LMG-collective_and_independent-decay-Lee} 

\begin{equation}
\label{eq:Master-gleichung}
\dot{\hat\rho} = -i [\hat{H},\hat\rho] - \frac{\kappa}{2N} \mathcal{D}[\hat{J}^+]\hat\rho, 
\end{equation}
where $\mathcal{D}[\hat{J}^+]\hat{\rho} = \hat{J}^-\hat{J}^+ \hat{\rho} + \hat{\rho} \hat{J}^-\hat{J}^+ - 2 \hat{J}^+ \rho \hat{J}^-$ is the Lindblad-Dissipator and $\hat{J}^\pm = \hat{J}_x \pm i \hat{J}_y$. \\
To compensate the dissipative effects we assume a time dependent coupling $\gamma_x$ of Pyragas form \cite{pyragas1992continuous} . Therefore we condition $\gamma_x$ to depend on a difference of $\hat{J}_z$ averages at two different times with time delay $\tau$ as following  
\begin{equation}
\label{eq:Kopplung_with_Pyragas}
\gamma_x = \gamma_x(t) = \gamma + \frac{\lambda}{N^2} (\avg{\hat{J}_z(t-\tau)}^2 - \avg{\hat{J}_z(t)}^2), 
\end{equation} 

In the following, we investigate the ESQPT signal at the quantum level using the effective Hamiltonian approach and at a semiclassical level using the solution of mean-field equations in thermodynamic limit. In the second case we show the unique effect of the feedback loop especially in context of the ESQPT signal.

\section{Dissipative ESQPT at quantum level}
The fate of the ESQPT signal in presence of dissipation is still not studied well. In closed systems the ESQPT is hidden in the energy spectrum or is visible in observable averages as a function of energy \cite{ESQPT_in_quantum_optical_models-Pedro} as a non-analyticity, which can be obtained by quantum mechanical or semi-classical calculations. Its origin can be understood at a semi-classical level as critical points from the energy surface. 

\begin{figure}[t]
\centering{\includegraphics[width=1\columnwidth]{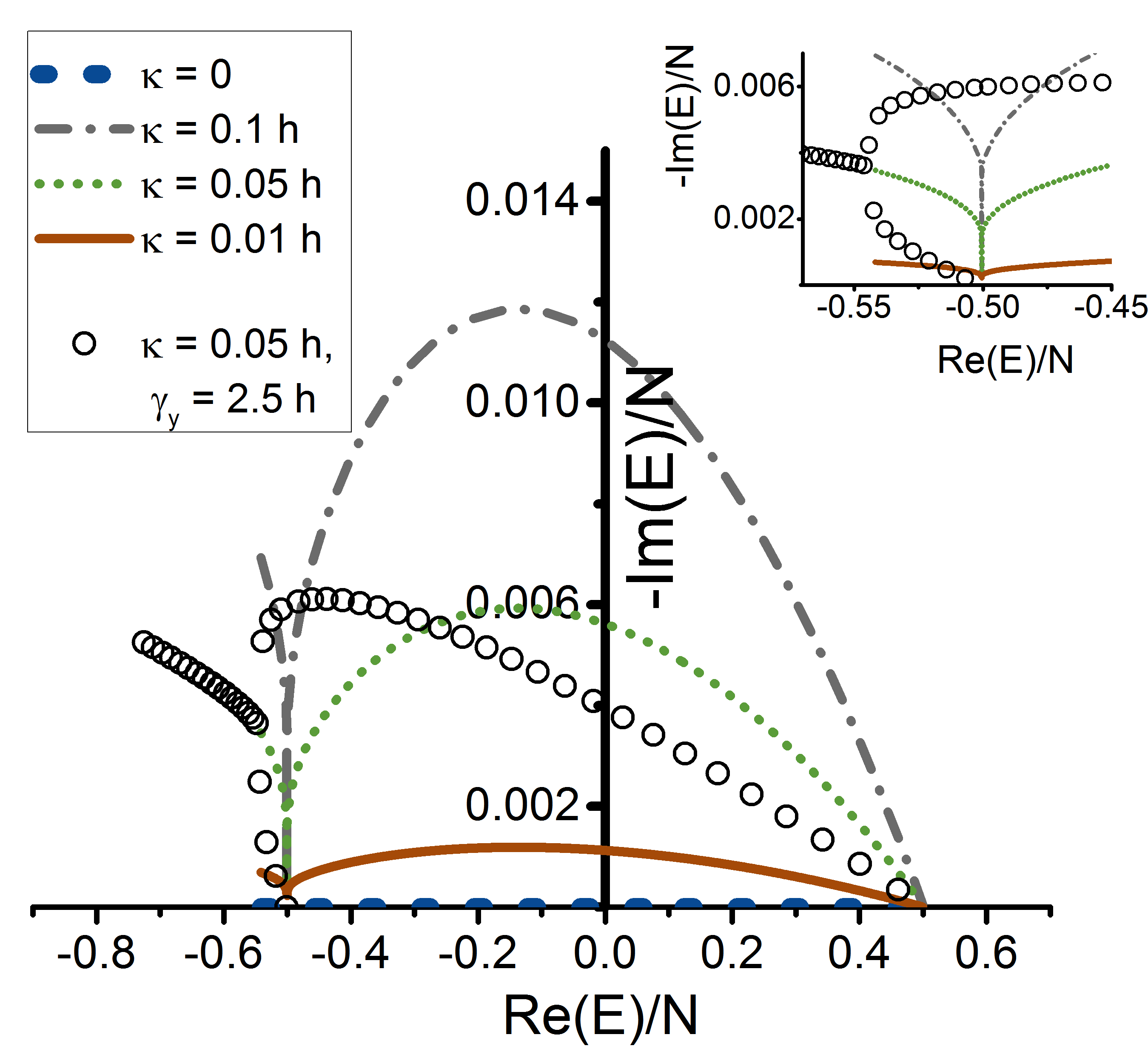}
\includegraphics[width=1\columnwidth]{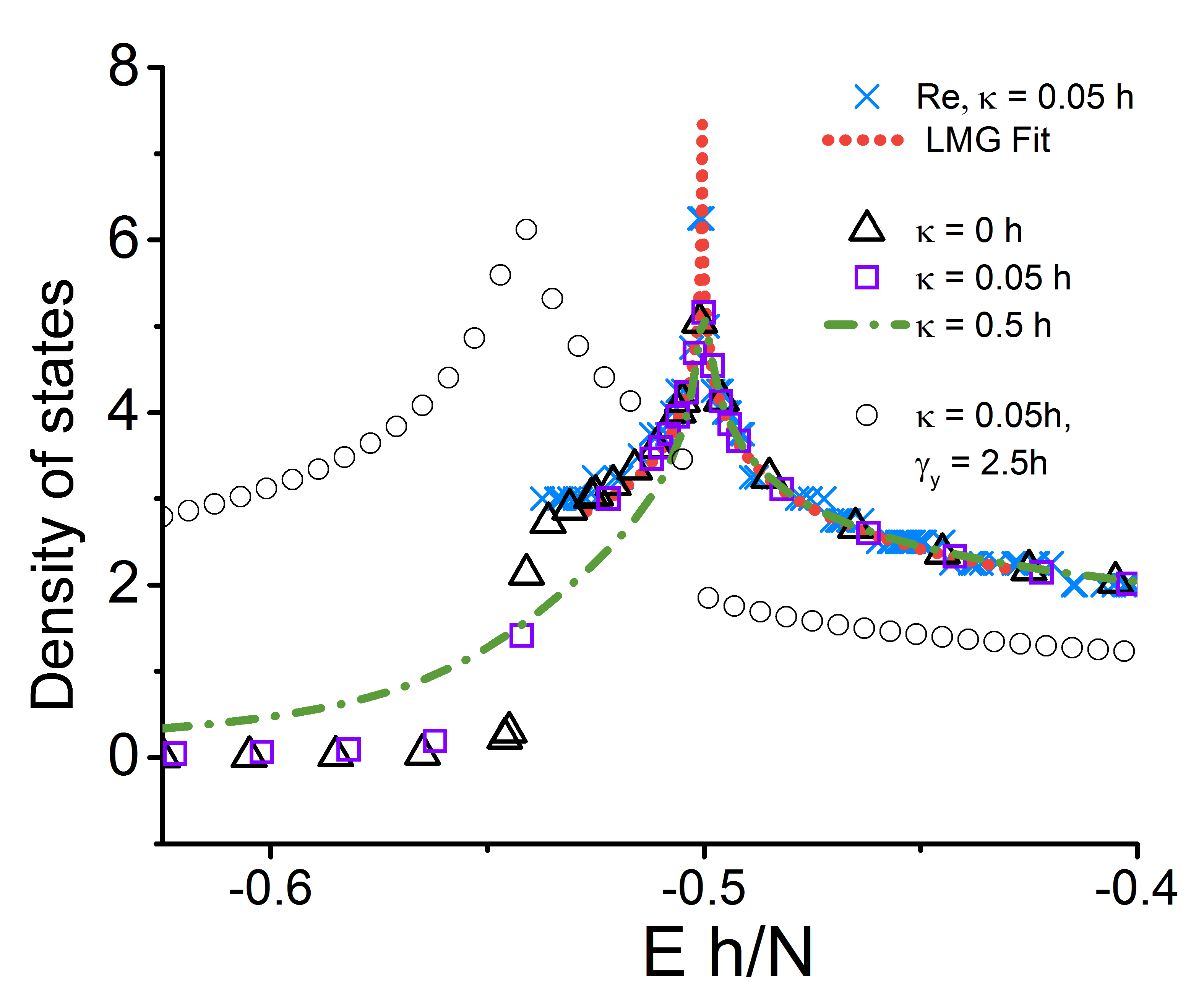}
}
\caption[]{\label{fig0} (upper) The complex spectrum for different values of $\gamma$. The $\curlyvee$-structure indicates the position of the ESQPT. The inset shows the zoom around the ESQPT energy.  (lower) The density of states of the effective LMG Hamiltonian shows a logarithmic divergence at $E = -0.5 N$ . For comparison, the $\gamma_y \neq 0 $ case is additionally shown in both figures (open circles). Parameters: $(\gamma=1.5h),N=1000-10000$,$\lambda=0$.
}
\end{figure}

In dissipative systems described by a Lindblad master equation, an effective non-hermitian Hamiltonian can be defined in a standard way. Rewriting the master equation Eq. \eqref{eq:Master-gleichung} as 
\begin{equation}
\dot{\rho} = -i [H_{eff},\rho] + \frac{\kappa}{N} J^+ \rho J^-,
\end{equation}
we obtain the effective non-hermitian Hamiltonian 
\begin{equation}
\label{eq:Heff}
H_{eff} = H - i \frac{\kappa}{2N} J^- J^+
\end{equation}

with complex spectrum, which is shown in the Fig. \ref{fig0} (upper) for different values of $\kappa$ in the symmetry broken phase. For $\kappa \neq 0$ the spectrum has an imaginary part which scales with $\kappa$. The imaginary part can be interpreted as a decay rate at a certain energy level \cite{Jung-Phase_transition_open_qs,Rotter-spectr-properties_of_excited_states}. To complete the spectral information we show in the lower part of Fig. \ref{fig0} the corresponding density of states. For example, open triangles show the density of states along the blue dashed line in the upper figure in the known $\kappa = 0$-case \cite{LMG-spectrum_thermodynamic_limit_and_finite_size-corr-Mosseri}, where a logarithmic divergence at $E = -0.5 N$ is due to the ESQPT.      

How does the dissipation affect the  ESQPT? The ESQPT survives and, somewhat surprisingly, it becomes visible not only in the density of states of the non-hermitian Hamiltonian $H_{eff}$, but can be also directly seen from its complex eigenvalues, see Fig. \ref{fig0} (upper). The ESQPT is hidden in this representation in a $\curlyvee$-form at the energy of $E = -0.5 N$. Another feature (due to the assumed Lindblad operator \eqref{eq:Master-gleichung}) is the vanishing of the imaginary part of $H_{eff}$ at the north and south poles of the corresponding Bloch sphere, which leads to zero imaginary part at the corresponding energies of $\mp 0.5 N$. Thus the decay rate is zero there and the dissipative effects disappear at this points. Note, this effect is also present at the level of the mean-field Eqs. \eqref{eq:semiclassical_eq}, where dissipative terms do not contribute at the poles for arbitrary $\kappa$ values due to conservation of the spin length.  

We used two different methods to calculate the density of states in the $\kappa = 0.05 h$ case. First, we used only the real part of the eigenvalues and counted their number in a certain energetic window (blue crosses in Fig. \ref{fig0}). Second, we used 
\begin{equation}\label{eq:density_of_states_pfadintegral}
\nu(E) = -1/\pi \Im \text{Tr} \left(\frac{1}{E - H_{eff}}\right)
\end{equation} 
with the non-hermitian Hamiltonian (open squares). We emphasize, that the results agrees and have still a logarithmic divergence at the energy $E = -0.5 N$ which can be well   fitted by a log-function (red dotted curve). Especially for $E > -0.5N$ there is no strong deviation from the $\kappa = 0$ case (open triangles). Only for $E< -0.5N$, there is a deviation from the non-dissipative result, which can be better seen for the curve with a bigger dissipation rate $\kappa = 0.5$ (green dot-dashed).    

We emphasize, that only the $\curlyvee$-structure and not the zero imaginary part in the complex spectrum indicates the ESQPT. For $\gamma_y = 0$ both effects are at the energy $E=-0.5$, though they can be easily separated for an anisotropic LMG Hamiltonian $(\gamma_y \neq 0)$, where the ESQPT can be shifted to other energies and a jump in a density of states can occur on top \cite{LMG-spectrum_thermodynamic_limit_and_finite_size-corr-Mosseri}. Setting $\gamma_y = 2.5h$, we observe a shift of the $\curlyvee$-signal to the corresponding energy of ESQPT in this case (open circles in the upper part of Fig. \ref{fig0} ). The corresponding density of states (open circles in the lower part of the figure) has a peak at the ESQPT energy $E\approx -0.54 N$ and a jump at $E = -0.5 N$ \cite{LMG-spectrum_thermodynamic_limit_and_finite_size-corr-Mosseri}. Note, that the finite size effects smooths the peak and the jump in this case.



\section{Dissipative ESQPT signal  at mean-field level}

\subsection{Mean-field-equations}
To derive a mean-field equation for the dissipative LMG system, we use $\avg{\hat{O}} = Tr(\hat{O} \hat{\rho})$, assume the factorisation assumptions $\avg{\hat{O}_1 \hat{O}_2} \approx \avg{\hat{O}_1} \avg{\hat{O}_2}$  which is known to hold in the thermodynamic limit and to forecast the same observable averages and the phase transition as the quantum mechanical calculations \cite{LMG-Finite_size_scalling_Dusuel,LMG-TC-periodic_dynamic_and_QPT-Georg}. We then obtain a following set of closed semiclassical equations of motion \cite{Morrison-Dissipative_LMG-and_QPT}

\begin{align}
\label{eq:semiclassical_eq}
\dot{J}_x(t) &= h J_y(t) -\kappa  J_x(t) J_z(t), \\
\dot{J}_y(t) &=-h J_x(t) +2 \gamma_x(t) J_x(t) J_z(t)-\kappa J_y(t) J_z(t), \notag\\
\dot{J}_t(t) &=-2 \gamma_x(t) J_x(t) J_y(t) +\kappa  \left(J_x(t)^2+J_y(t)^2\right), \notag
\end{align}
with rescaled averages $J_i = \frac{1}{N} \avg{\hat{J}_i}$. Without time delay $(\lambda = 0)$ the QPT, which is one important property in this system,  corresponds at this semi-classical level to a pitchfork bifurcation  \cite{Morrison-Collective_spin_system-QPT_and_entaglement}: Eq.\eqref{eq:semiclassical_eq} has two stationary states $J_{i}^0$, corresponding to a normal phase with $(J_x^0, J_y^0, J_z^0) = (0,0,1/2)$ and a symmetry broken phase 

\begin{align}
\label{eq:steady_state_S}
J_x^0 &= \pm \sqrt{\frac{-4h^2(\gamma - \sqrt{\gamma^2-\kappa^2}) + \kappa^2 (\gamma + \sqrt{\gamma^2-\kappa^2})}
				{8 \gamma \kappa ^2}}, \\
J_y^0 &= \frac{\gamma - \sqrt{\gamma^2-\kappa^2}}{\kappa} \cdot J_{x}^0 , 
J_z^0 = h \cdot \frac{\gamma - \sqrt{\gamma^2-\kappa^2}}{\kappa^2}, \notag
\end{align} 
whose stability swaps at a critical coupling

\begin{equation}
\gamma_{c} = h + \frac{\kappa^2}{4h}.
\end{equation}
Note, that even the dissipative model still fulfils the conservation law $J_x^2 + J_y^2 + J_z^2 = 1/4$, thus the dynamical evolution is restricted to a sphere. Furthermore, even with time delay ($\lambda > 0$) the fixed points remain the same, as the Pyragas term vanishes in the steady state. 

In the following, we investigate the ESQPT signal in presence of damping with and without control. Later we show that the time delayed  coupling $\gamma(t)$ may affect the linear stability of fixed points and the dynamical evolution of the system in a completely unexpected way, particular in its acting against the dissipation.

\subsection{Dissipative damping of the ESQPT}

\begin{figure}[t]
\centering{\includegraphics[width=1\columnwidth]{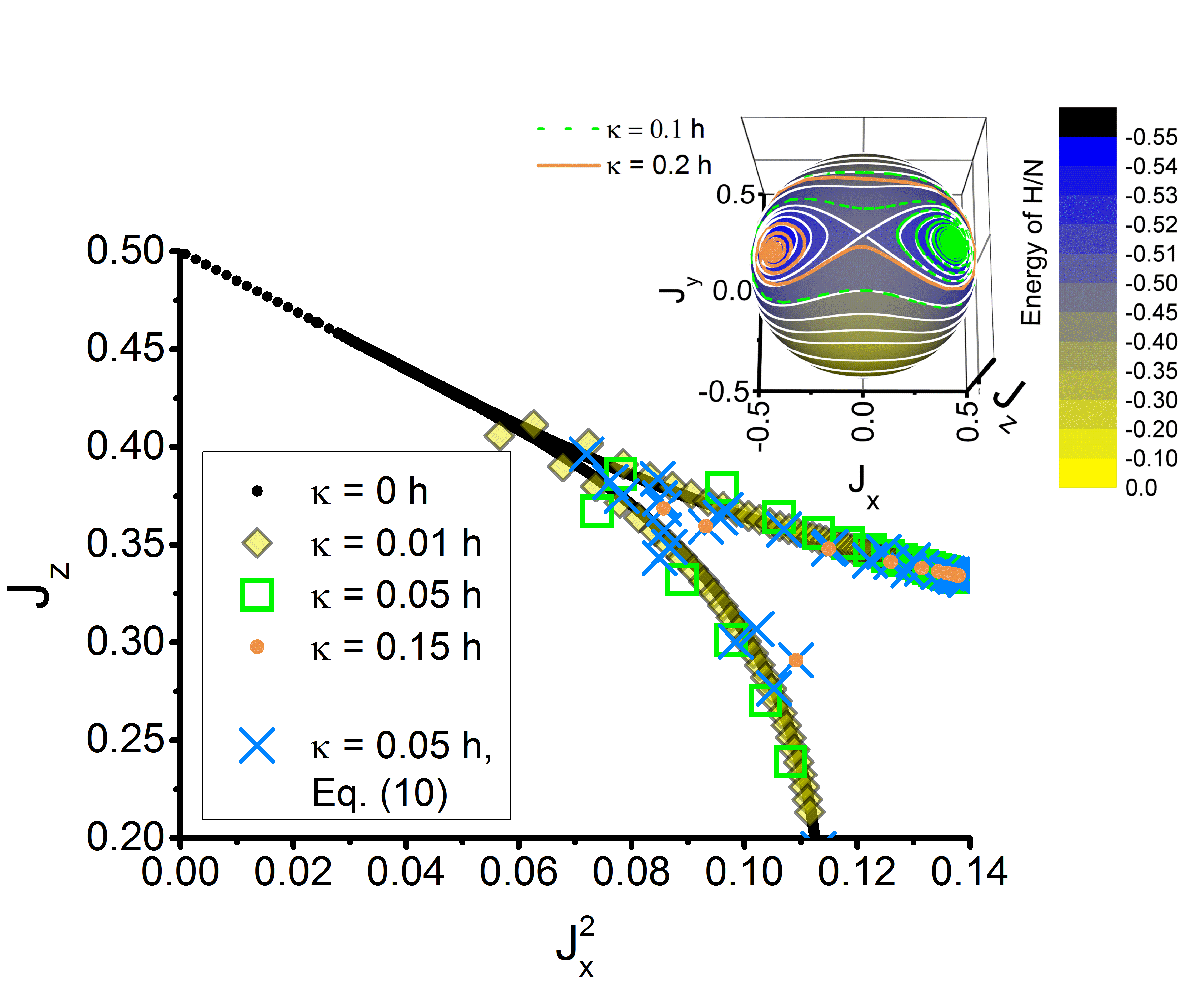}}
\caption[]{\label{fig2}The figure shows the smoothing impact of damping on the ESQPT signal for different $\kappa$ values.  The (blue) crosses show the averaging results using the integration method for $\Delta t = 20$. The ESQPT-signal under influence of damping gets smoothed. (Inset) Trajectories on the Bloch sphere visualize the system evolution Eq. \eqref{eq:semiclassical_eq} for $\kappa = 0.1 h$ (dashed green), $\kappa = 0.2 h$ (orange) and $\kappa = 0$ (white contours). The color shows the rescaled energy in $\kappa = 0$  case. Parameters: $\gamma=1.5h,\lambda=0$.
}
\end{figure}

Using the semiclassical equation of motion Eq. \eqref{eq:semiclassical_eq} we now study the action of dissipation on the ESQPT at this level. Therefore we first look at a dynamical evolution of the LMG system. The spin averages of the system are restricted to the Bloch sphere, which is shown in Fig. 2 (inset).  The color represents the energy for a given system configuration for $\kappa = 0$ and the white lines represent the paths with the same energy in the symmetry broken phase. Without dissipation the system follows one of this paths keeping the energy fixed, i.e. staying in the eigenstate. For each eigenstate one can compute an averaged value for $(J_x^2,J_z)$.
The black (dotted) curve (Fig. \ref{fig2}) shows this values for multiple eigenstates. This is a novel representation of two observable averages, which was recently suggested to visualize the energy independent ESQPT signal \cite{LMG-TC-periodic_dynamic_and_QPT-Georg}, which is visible again as a peak.    The continuation of the peak would end by $(0,1/2)$. In the upper half of a Bloch sphere we see a separatrix, a (white) path which goes through a north pole of the sphere. Due to the symmetry of the path, its averaged values for $J_x^2$ and $J_y^2$ should be zero, whereas the $J_z$ average is 1/2. Thus, paths close to the separatrix are responsible for the ESQPT peak in $(J_x^2,J_z)$ diagram. Without dissipation $(\kappa = 0)$ the period length for fixed energy (Fig. \ref{fig2}, upper right) diverges at the separatrix energy \cite{ESQPT-inmany_body_systems-Caprio}. 

For $\kappa \neq 0$ the energy is not conserved any more and the system tends oscillating around the $J_z$-axis to a steady state Eq. \eqref{eq:steady_state_S}. Fig. 2 (inset) shows two examples of the system state evolution $(J_x,J_y,J_z)(t)$ for two different $\kappa$ values, which where obtained by solving Eq.  \eqref{eq:semiclassical_eq} numerically. The ESQPT signal is now hidden in the dynamical evolution of the system. But, as we can see in Fig. \ref{fig2}, especially for big $\kappa$ values there are only less paths, which are close to the separatrix, thus the ESQPT signal will be damped especially for big dissipation rates $\kappa$. 
The impact of damping to the ESQPT signal in the $J_z,J_x^2$ plane can be obtained by calculating the mean values of $J_x^2(t), J_z(t)$-evolution for a period or by finding some optimal effective period $\Delta t$, in a way that the mean values calculated using the definition 
\begin{equation}
\label{eq:avarage_effectiv_period}
 \bar{O}(t) = \frac{1}{\Delta t}  \int_{t=a}^{a+\Delta t}  O(t) dt, \quad 
 a \in \{0,T_{max}\},
\end{equation} 
matches to the closed case as good as possible. Note, as a period length in the first case we take the time for one full rotation around the $J_z$-axis, which changes with time.   \\
In the first case, the results for different damping rates $\kappa$ are shown in Fig. \ref{fig2} by colored circles/diamonds and unfilled squares. We see, that the peak is now smoothed, but still visible especially for small damping rates. The blue crosses show the second case with effective period which lead to a much better ESQPT signal.

\section{Pyragas Control of the ESQPT signal}

Now we set $\lambda \neq 0$ in $\gamma_x(t)$ Eq. \eqref{eq:Kopplung_with_Pyragas} and investigate the impact of time delayed control on the system.  

\subsection{Linear stability analysis with time delay}
A usual approach to analyse the effects of time delayed feedback is first to check the stability of fixed points in the presence of control \cite{Schoell-Control_of_unst_states_by_time_del}. Therefore we linearise Eq. \eqref{eq:semiclassical_eq} around the fixed points $J_i(t) = J_i^0+\delta J_i$, obtaining the following system of linearised equations with $\delta  {\bf v} \equiv (\delta J_x, \delta J_y)^T$,

\begin{equation}
\label{eq:lin_eq}
\partial_t \delta {\bf v}(t) = {\bf B} \cdot \delta {\bf v}(t) + {\bf A} \cdot \delta {\bf v}(t- \tau),
\end{equation}
with 
\begin{widetext}
\begin{equation}
{\bf A} \equiv 
-4 \lambda  J_z^0 \begin{pmatrix}
 0 & 0 \\
 {(J_x^0)}^2   &  J_x^0 J_y^0 \\
\end{pmatrix}
 ,
{\bf B} \equiv
\begin{pmatrix}
 \kappa \frac{{(J_x^0)}^2}{J_z^0}- \kappa J_z^0  & h+ \kappa \frac{J_x^0 J_y^0 }{J_z^0} \\
 4 \lambda  J_z^0  {(J_x^0)}^2-2 \gamma \frac{ {(J_x^0)}^2}{J_z^0}+\kappa \frac{J_y^0   J_x^0}{J_z^0}-h+2\gamma J_z^0  &\kappa \frac{  {(J_y^0)}^2}{J_z^0}+4\lambda  J_x^0 J_z^0  J_y^0- 2\gamma \frac{J_x^0 J_y^0}{J_z^0}- \kappa J_z^0  \\
\end{pmatrix},
\end{equation}
\end{widetext}

with eliminated $J_z$ component by use of the spin-length conservation law. 

The roots of the corresponding characteristic equation 

\begin{equation}
\label{eq:stability_eq}
\det(\Lambda {\bf 1} - {\bf B} - { \bf A} \exp(-\Lambda \tau)) = 0
\end{equation} 
determines the stability of a fixed point, which is stable if all real parts of all solutions $\Lambda$ are negative \cite{schoell-Handbook_of_chaos_control}.

\begin{figure}[t]
\centering{\includegraphics[width=0.99\columnwidth]{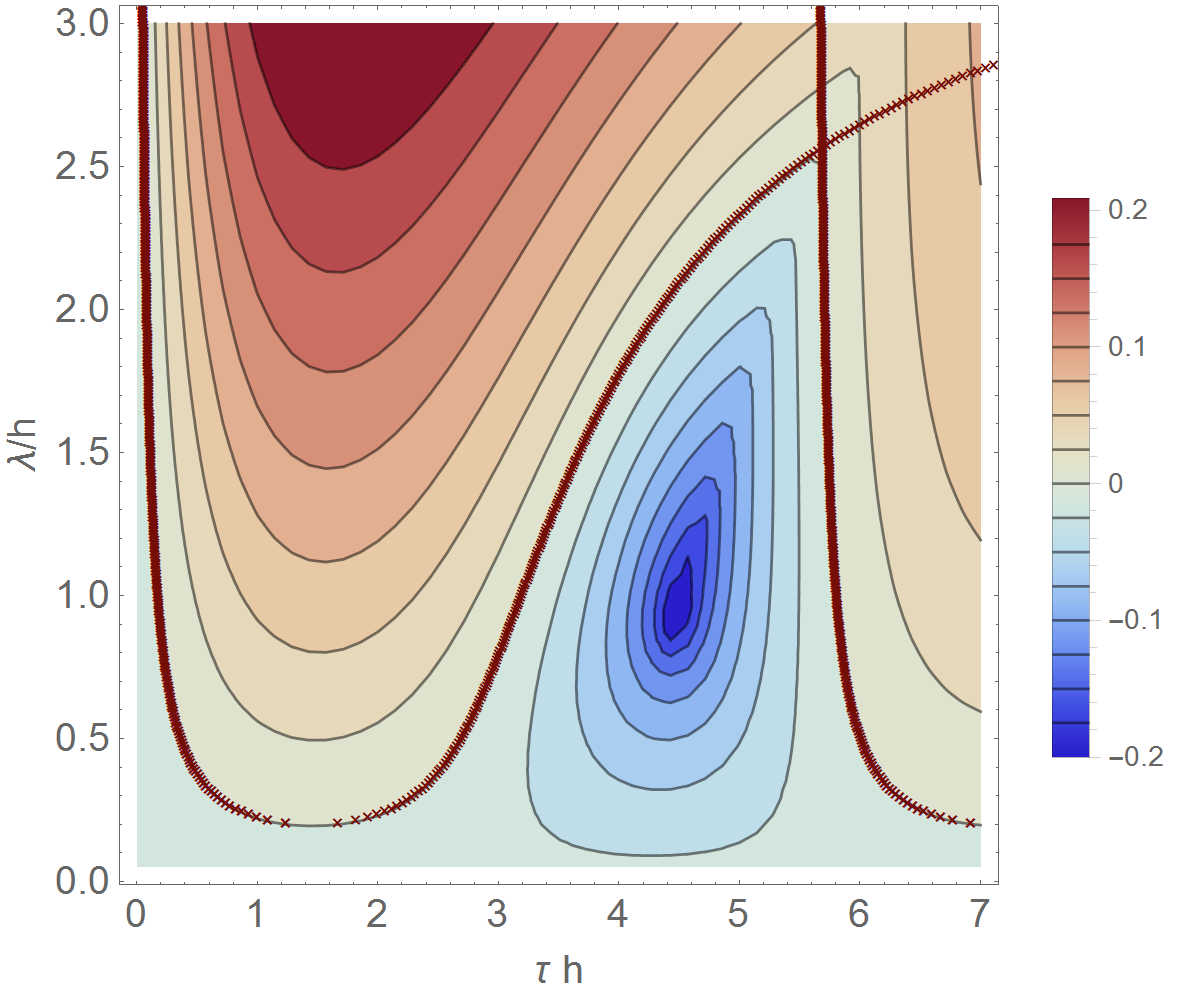}
}
\caption[]{\label{fig1} Stability diagram in the $\tau-\lambda$-plane. The color represents the stability robustness of fixed points in the symmetry broken phase, which are not stable for positive values (red coloured area). The brown line defines a boundary condition and has an analytical expression.  Parameters: $\gamma=1.5h,\kappa=0.05 h$.
}
\end{figure}

In Fig. \ref{fig1} we plot the biggest real part of eigenvalues in the $\tau-\lambda -$ domain. We see that there is a window for $\lambda$ -values, there the stability of fixed points oscillates from stable to unstable and from unstable to stable while increasing the time delay. Outside this window, the fixed points remains either stable $(\lambda \ll 1 h)$ or lose their stability forever $(\lambda \gtrsim 2.5 h)$. Note, that the boundaries between stable and unstable zones (brown line) can be calculated from an analytical expression, see Appendix \ref{ap:boundaries}. Next, we analyse the system properties in the unstable regime, use them to obtain a sharp  ESQPT signal and show chaotic behaviour for larger time delays $\tau$.

\subsection{Feedback compensates dissipation}

\begin{figure*}[ht]
\centering{\includegraphics[width=0.24\columnwidth]{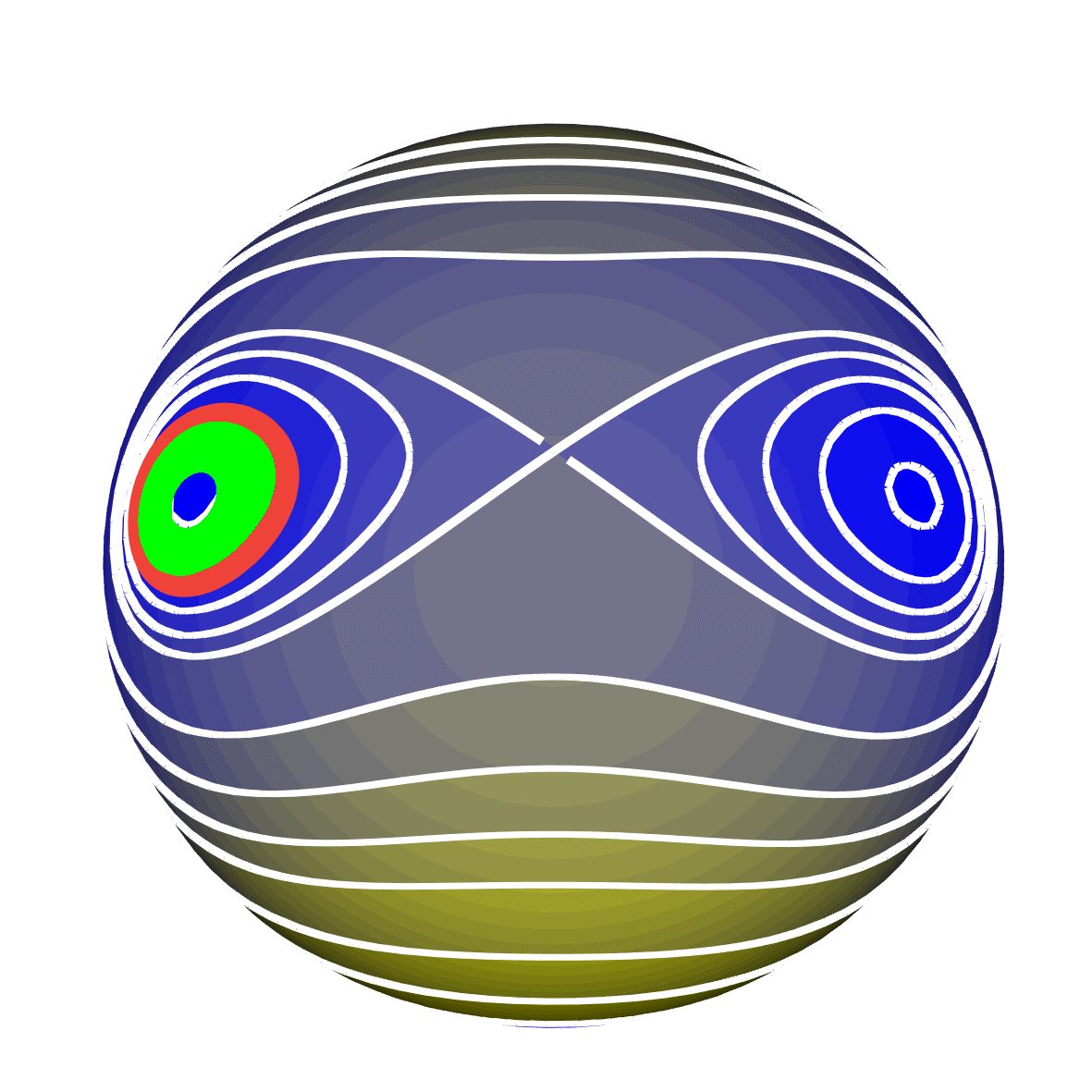}
\includegraphics[width=0.24\columnwidth]{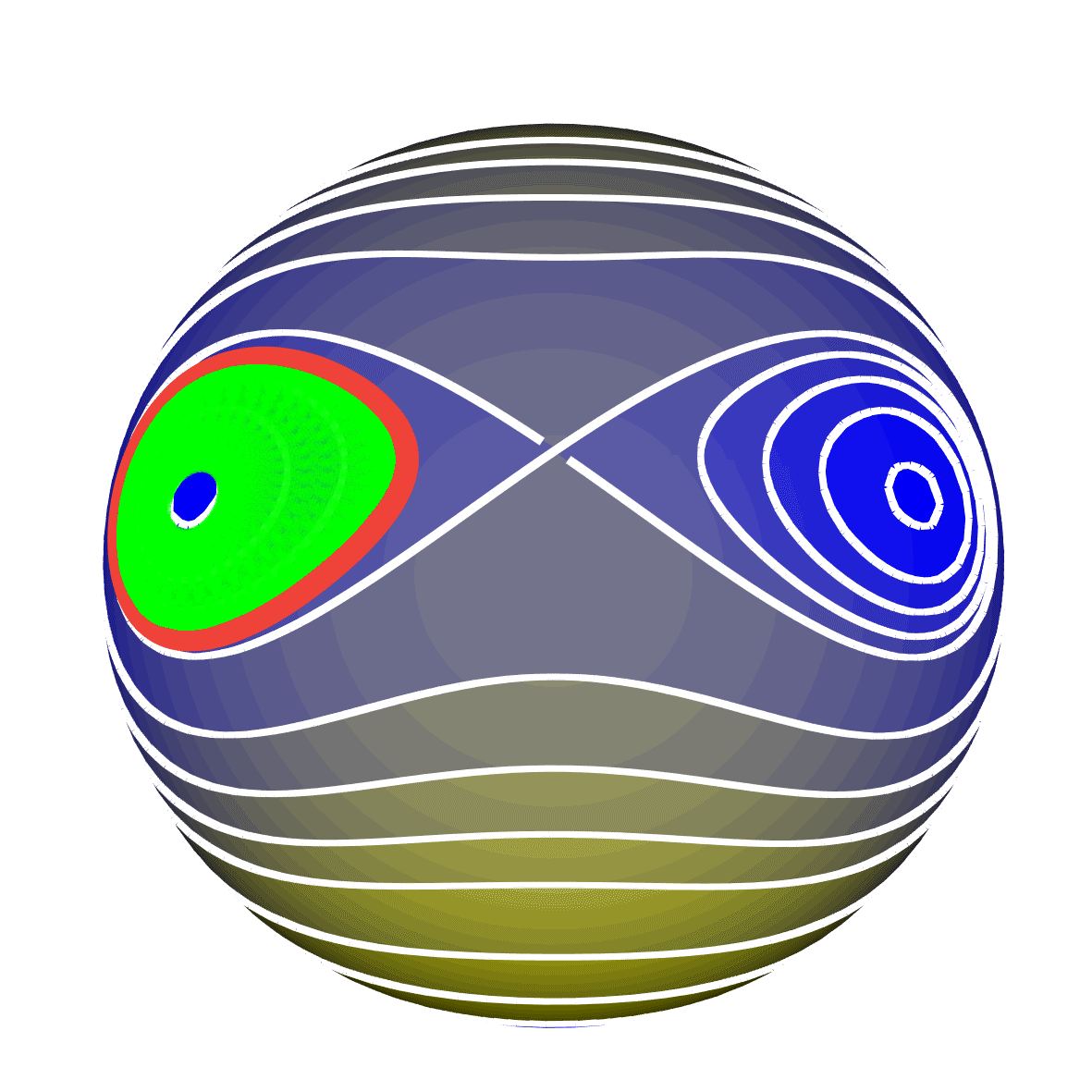}
\includegraphics[width=0.24\columnwidth]{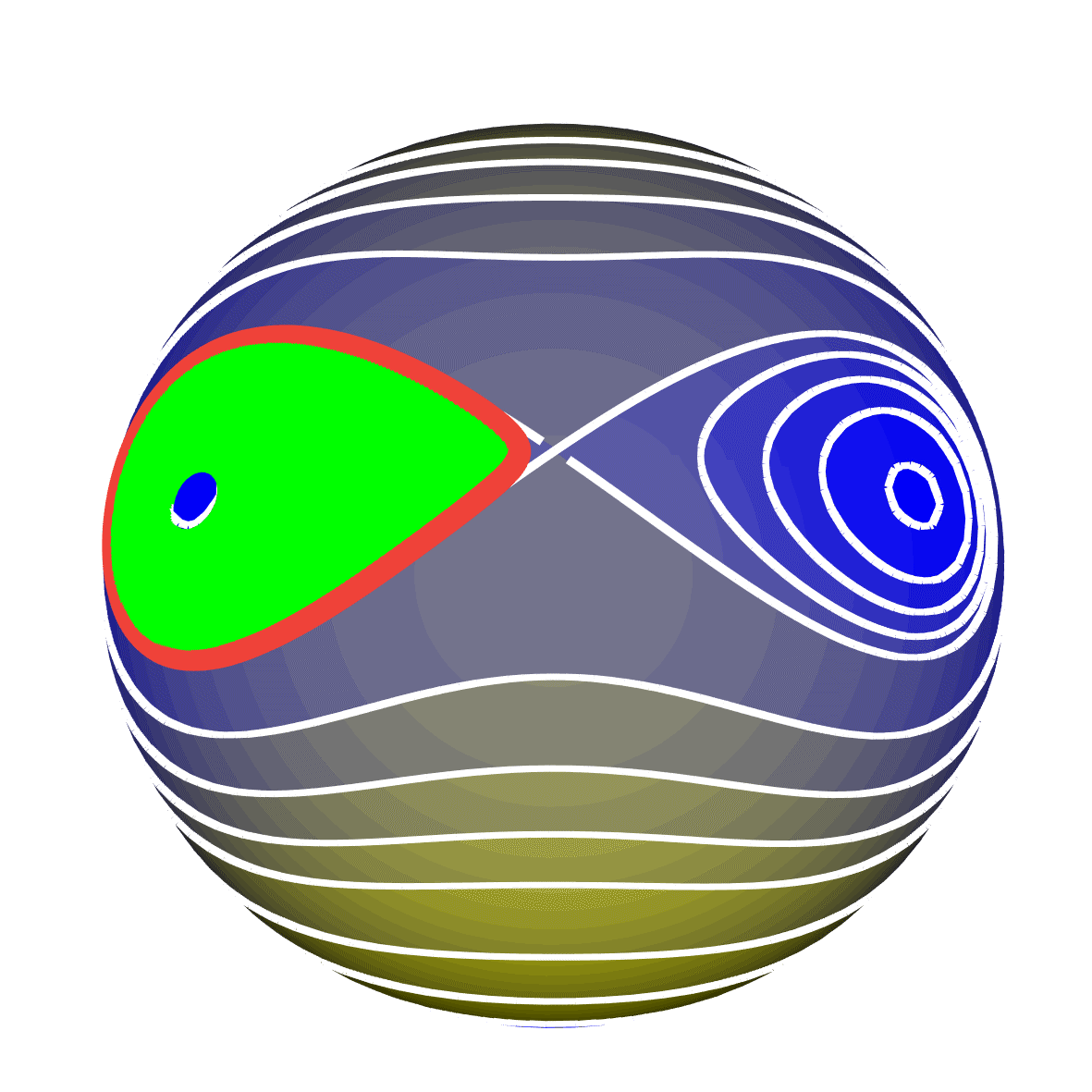}
\includegraphics[width=0.24\columnwidth]{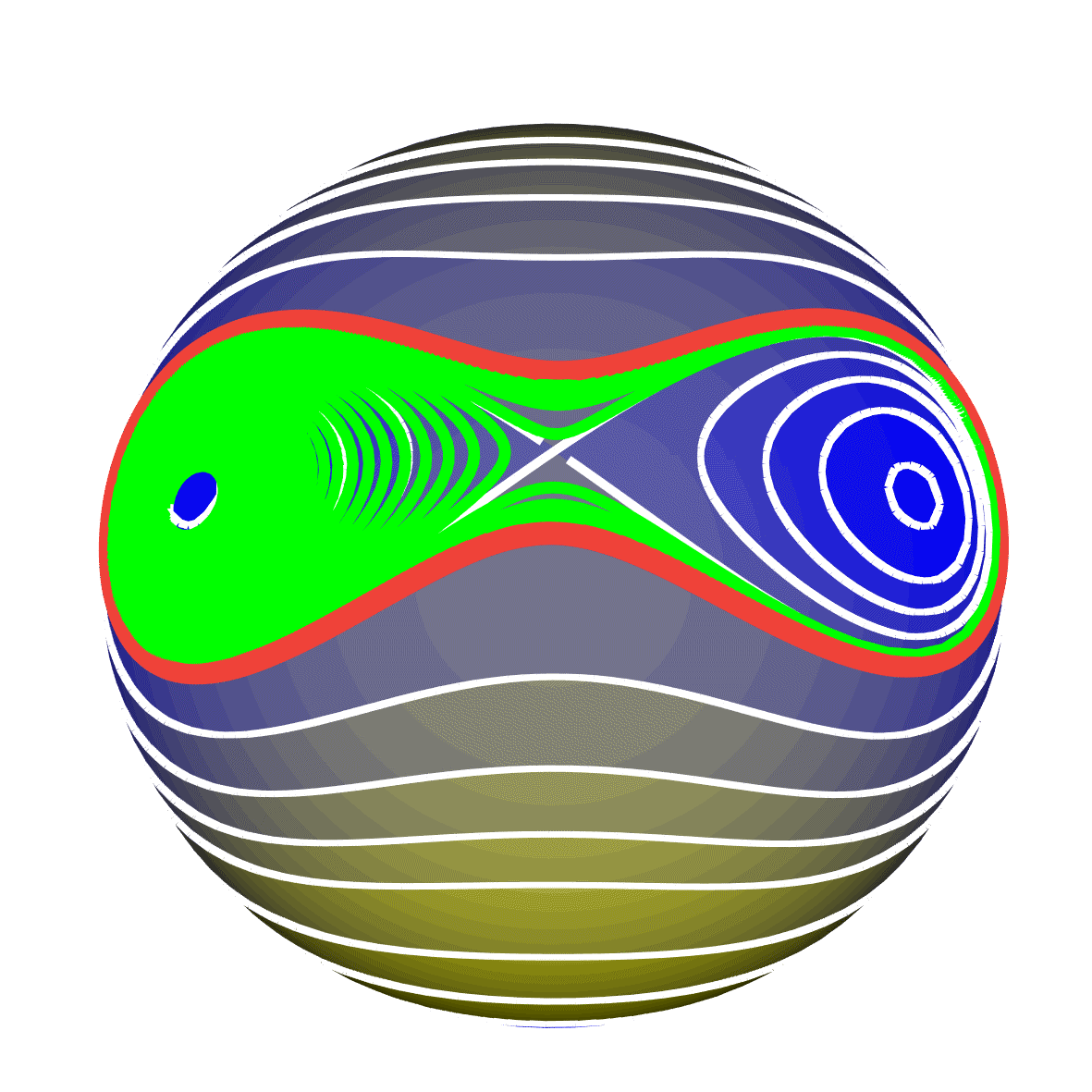}
\includegraphics[width=0.24\columnwidth]{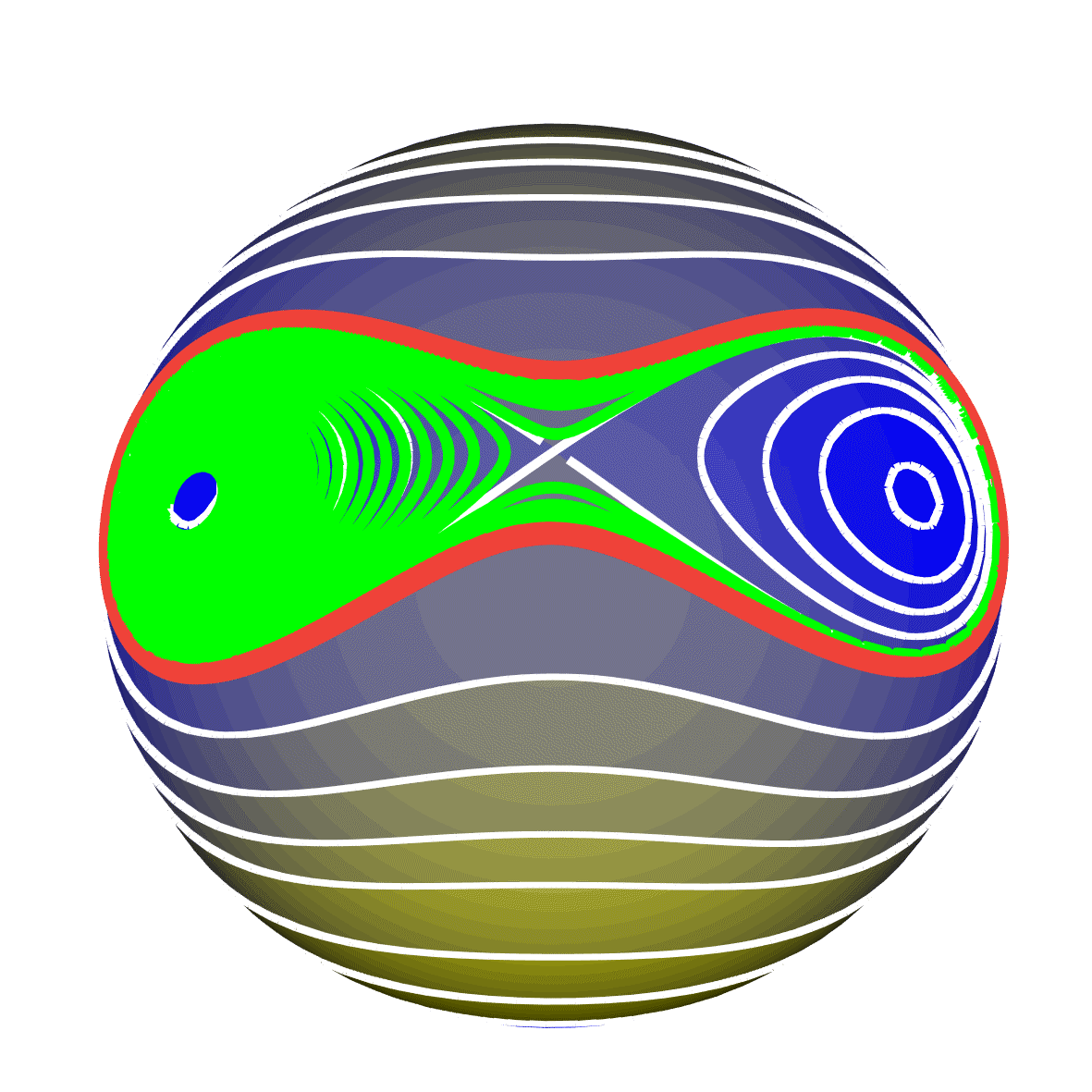}
\includegraphics[width=0.24\columnwidth]{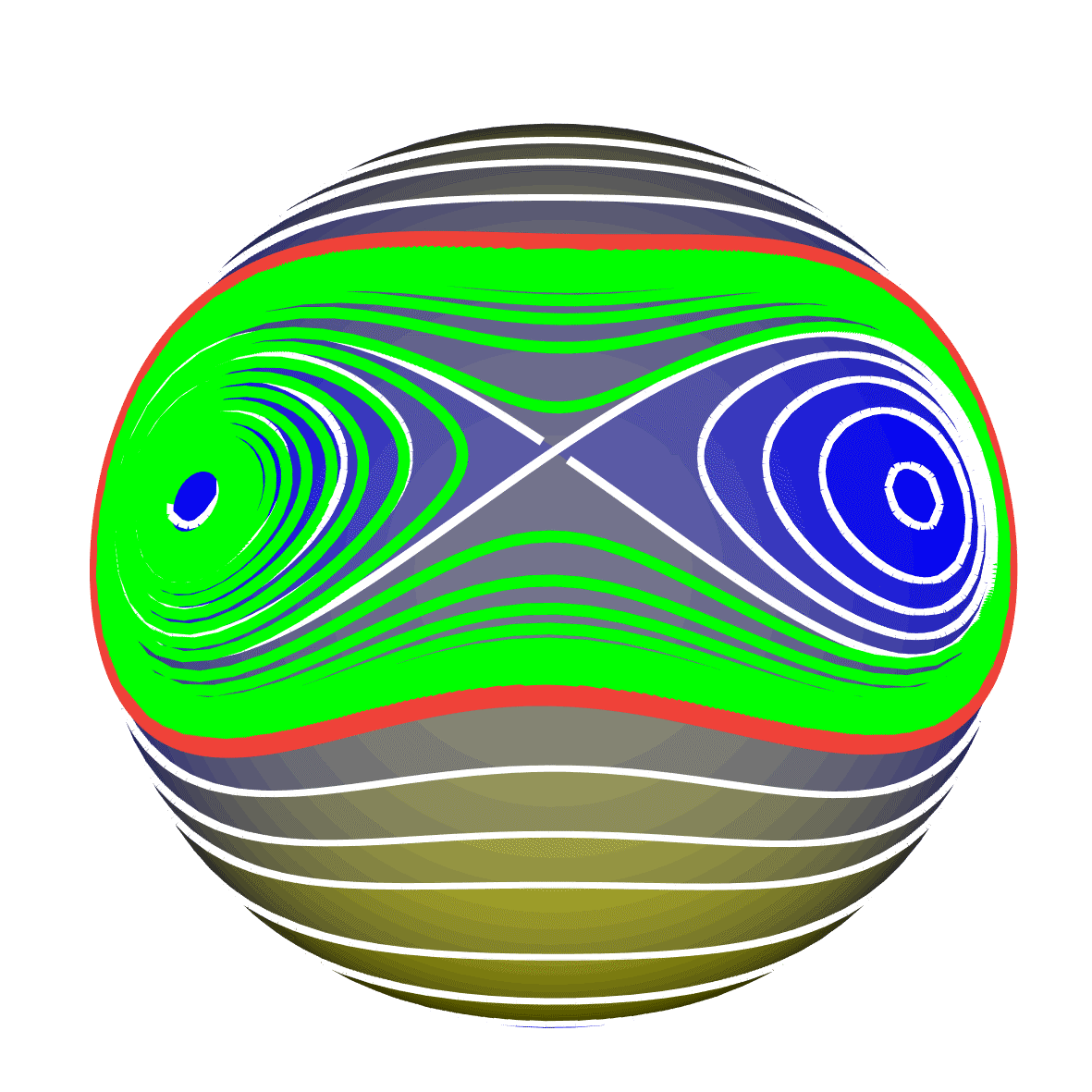}
\includegraphics[width=0.24\columnwidth]{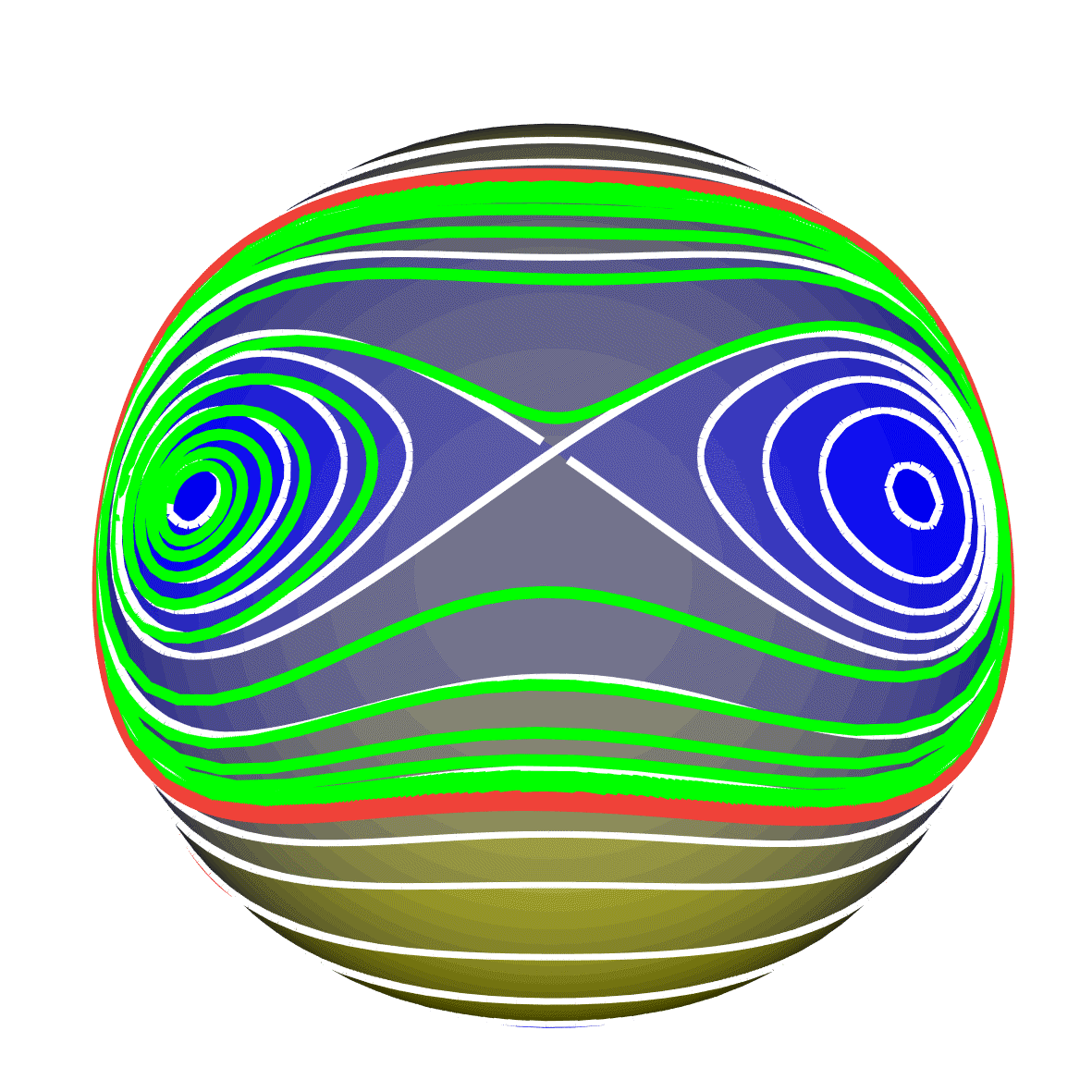}
\includegraphics[width=0.24\columnwidth]{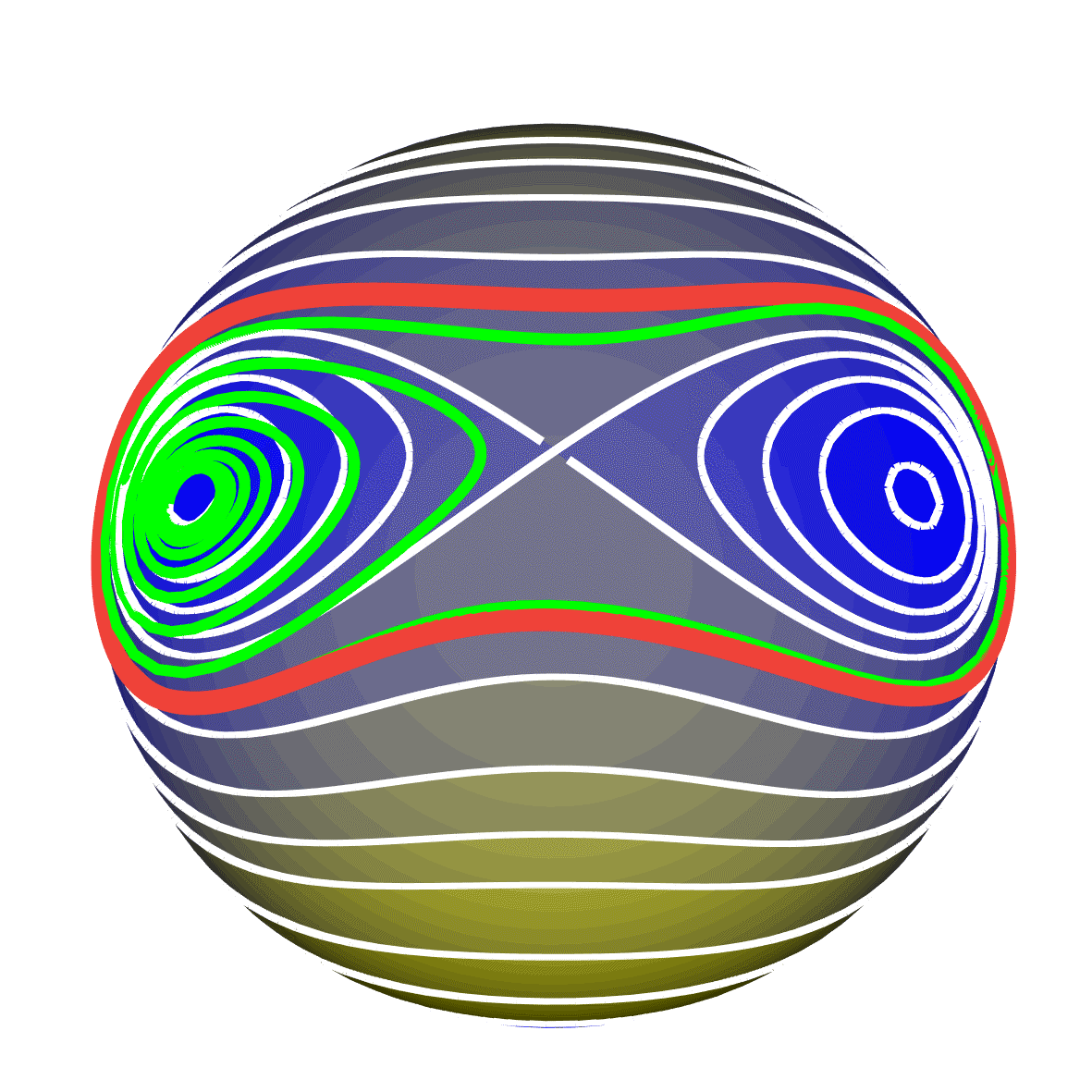}
}
\caption[]{\label{fig3} Dynamical evolution of the controlled LMG system on the Bloch sphere for the following $\tau$-values (from left to right): $\tau h = 0.2; 0.25; 0.3; 0.31; 0.35; 0.5; 1;  2.5; $ The red thick path represents the stationary state. Increase of $\tau$ forces the solution to cross the separatrix. In this way the ESQPT signal (Fig. \ref{fig4}) is restored from stationary solutions with different $\tau$ values. Note, the axes and scaling are as in the inset of Fig. \ref{fig2} but are not visible due to breakdown.  Parameters: $\gamma=1.5h,\lambda=1, \kappa = 0.05 h$. 
}
\end{figure*}

Our feedback scheme Eq. \eqref{eq:Kopplung_with_Pyragas} modifies the system dynamics in an interesting and unexpected way. Increasing the time delay $\tau$, we cross the boundary and make the fixed point unstable. For smaller $\tau$ values the trajectory ends in a new stable state in form of a limit cycle, thus a Hopf bifurcation occurs. Moreover the trajectory of the limit cycle has only small deviations from paths with fixed energy, which the LMG system would take without dissipation. The size of trajectories  can be changed by $\tau$, thus tuning the time delay value corresponds to a change of energy in a closed LMG system. Fig. \ref{fig3} demonstrates the feedback action, showing the trajectory evolution for different values of $\tau$.  Here we start close to the fixed point. The red (thick) curve shows the stationary state. Note, that the change of initial condition to the region outside the separatrix leads to the same stationary state for the considered $\tau$ values. \\

\begin{figure}[ht]
\includegraphics[width=0.99\columnwidth]{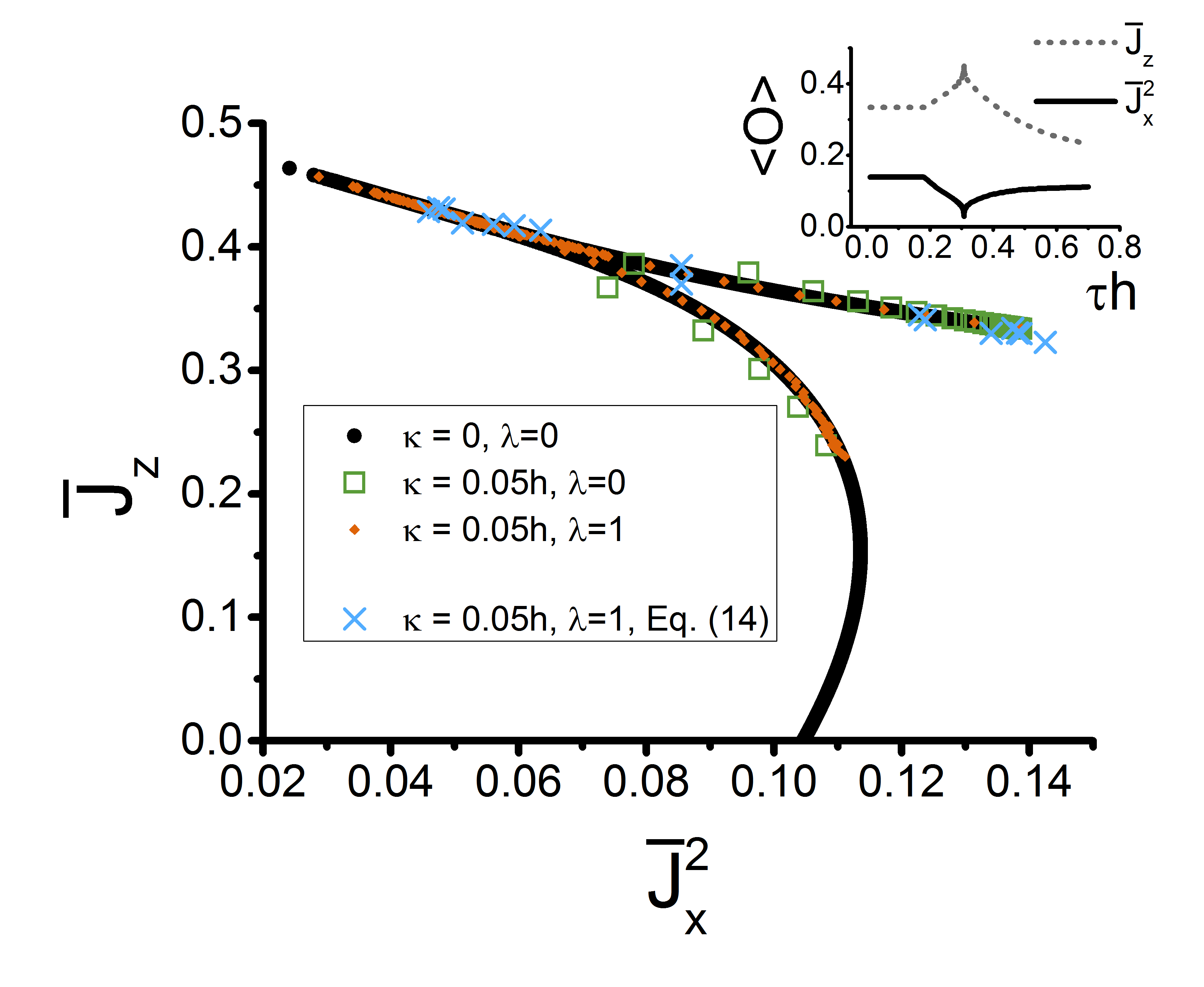}
\caption[]{\label{fig4} Time delayed feedback undo the dissipative effects to the ESQPT signal. The orange (light-coloured)rhombus shows averaged system values $\bar{J}_z,\bar{J}_x^2$ in the stationary state for fixed $\tau$ with $\tau \in [0.1,0.6]h$, which matches to the signal of the closed system (black curve) very well. Unfilled squares represent the ESQPT-signal under dissipation. The blue crosses shows the ESQPT-signal obtain from a dynamical evolution of the system for one big $\tau$-value using Eq. \eqref{eq:avarage_definition_in_chaos_fall}. (Inset) $\bar{J}_z,\bar{J}_x^2$ as a function of $\tau$ in a steady state.  Parameters: $\gamma=1.5h, \lambda = 1h$.
}
\end{figure}

Thus, using the stationary limit cycle states for fixed $\tau$ values offers a possibility to obtain an ESQPT-Signal again, by calculating the 
\begin{equation}
\label{eq:avarage_definition_in_chaos_fall}
\bar{J}_z = \frac{1}{t_2-t_1}\int_{t_1}^{t_2} J_z(t) dt 
\end{equation}
and $\bar{J}_x^2$ (with similar definition) averages  for fixed values of $\tau$ ($t_2 > t_1$ and  with $t_1$ big enough to become a stationary solution). Fig. \ref{fig4} shows the results of this calculation and compares them with the closed and dissipative cases without feedback. The (orange) rhombi shows the $\bar{J}_z,\bar{J}_x^2$ time averages in stationary limit cycle phase of the dissipative LMG system, the black curve shows the ESQPT -signal of a closed system without control. Each rhombus has its own time delay, the inset in Fig. \ref{fig4} shows the $\tau$-dependence of the averaged values. Note, that up to $\tau \approx 0.2/h$ the fixed point is stable and the averaged values are the values of the fixed point. We see a very good overlap between the control caused and the original signals, thus our feedback schema  compensates the dissipative smoothing (unfilled green squares) of the ESQPT-signal very well.

\subsection{Chaotic behaviour}

\begin{figure}[t]
\includegraphics[width=0.99\columnwidth]{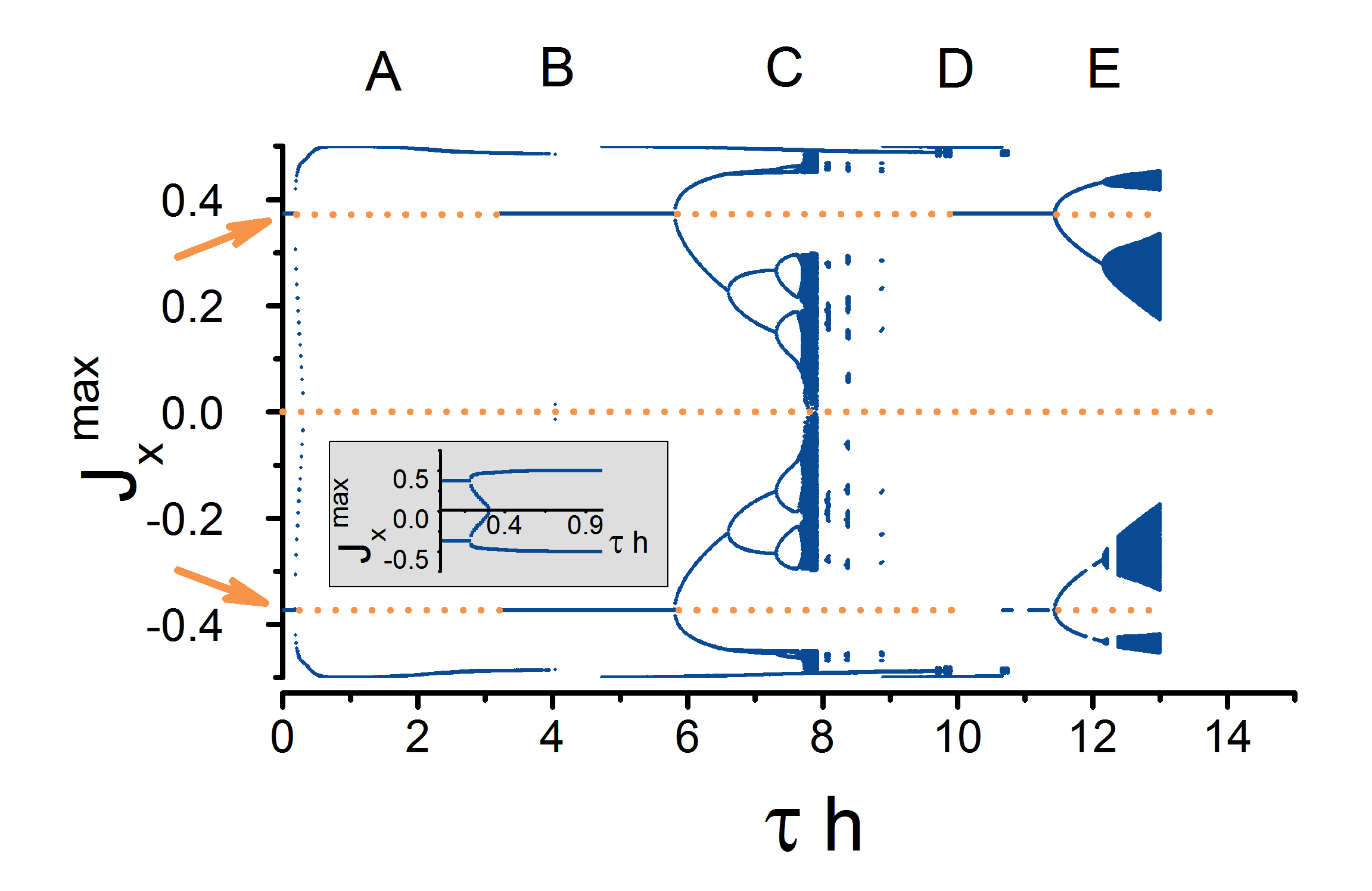}
\caption[]{\label{fig5} Bifurcation and period doubling for different time delays $\tau$. For each $\tau$-values the values of local maxima and minima of $J_x(t)$-value are plotted in stationary case. The dashed orange lines shows the unstable fixed points. The arrows points to the non-trivial fixed points. The phase diagram is divided the regions A-E with different properties. The inset shows shows a zoom of area A. Parameters: $\gamma=1.5h, \lambda=1h$.
}
\end{figure}

For $\tau \gg 1$ the stationary dynamics can become much more complex than just a creating limit cycles. Oscillations with more than one maximum and minimum appear, which is known as a way to chaos by period doubling \cite{Strogatz-nonliniear_dynamic_chaos,Laser-Bifurcation_Feedback-Hohl}. In Fig. \ref{fig5} (upper) we plot the maxima and minima of $J_x(t)$ oscillations for $t \gg 1$ as a function of time delay $\tau$ and mark different areas by capital letters A-E. In the inset we show a zoom for the area A. 
Increasing $\tau$ from zero, the stable fixed points (arrow) loses their stability (orange dotted line) at the first boundary condition from the Fig. \ref{fig1} and a Hopf-bifurcation appears (part A). In part B, the fixed points become stable again. But there exist still a stable limit cycle solution with rather big $J_x$-amplitude. Thus, in area B two stable solutions are possible, which is not in contradiction to the stability of the fixed point,  as the initial condition is not chosen in a way to fulfil the linearised assumption.  For $4 \lesssim \tau \lesssim 5 $ this limit cycle disappears.    
Further increase of time delay leads to a creation of a period doubling structure (area C), which is separated by windows where the solution converges to a limit cycle. In the region D the fixed point becomes stable again and the double period structure is gone, whereas in the Region E it appears again. Note, that the dotted line at $J_x^{\text{max}}=0$ represents the unstable trivial fixed point. 

Such chaotic behaviour can also be used to obtain the ESQPT signal, see blue crosses in Fig. \ref{fig4}. Fixing the time delay in the chaotic area of the phase diagram (Fig. \ref{fig5}), one can use the integration method with effective period Eq. \eqref{eq:avarage_effectiv_period} to obtain the shown curve from the dynamical system evolution on the Bloch sphere. We see, that also this result matches pretty well to the original ESQPT signal.

\section{Discussion}
In this paper we have demonstrated the effect of dissipation on the ESQPT signal for the LMG model and showed how to compensate it using time delayed Pyragas feedback modulating the interaction parameter between the atoms. Our results show, that the ESQPT is encoded in the spectral properties of the effective Hamiltonian as well as it is visible in the measured averaged values of the spin components. In the last case, smoothing effects appear which can be undone by our feedback scheme.  

We think, that an experimental measurement of an ESQPT signal using the time delayed method is easier than in an ideal, closed system. Using time delay one has only to measure the system values for different time delays $\tau$, instead of preparing the system in eigenstates or coherent superpositions to obtain the same information \cite{LMG-TC-periodic_dynamic_and_QPT-Georg}. 

We also checked the interaction strength $\gamma_x$ depending on other operator differences instead of $J_z^2$, or a modulated magnetic field $h$ instead of $\gamma_x$. However, in both cases the general dynamical properties remains the same. The LMG system has still limit cycles close to the separatrix for some fixed time delayed values and has a parameter range with chaotic behaviour. Both effects might be interesting from an experimental point of view. On the one hand, it is easier to control the magnetic field $h$, on the other hand, choosing another feedback loop can shift the appearing effects to other $\tau$ values, which could be easer to realize.  

We also tried to find ESQPT signatures in the correlation function \cite{Carmichael_stat-methods-in_Quantum_optic1} $C_t(z)\equiv \avg{\hat{J}^+(t+z)\hat{J}^-(t)}$, using the quantum regression theorem \cite{Schaller-QS_far_from_equilibrium}, but have not succeed, as the ESQPT signal (which should be visible as a peak at zero frequency in the Fourier space) interfere with effects like macroscopic occupation and degenerated spectrum in the symmetry broken phase. However one possible way out could be to drive the LMG system additionally with an external laser and calculating the resonance fluorescence spectrum, but this would require a re-derivation of the master equation. Note, that such spectra have been already calculated for an LMG system, but only in the linearised version \cite{Morrison-Collective_spin_system-QPT_and_entaglement}.

Similar to the Dicke model with feedback \cite{Kopylov-time_delayed_control_Dicke}, the Pyragas controlled LMG model has  stable limit cycle phases. In contrast to the Dicke model,  it shows chaotic dynamics for bigger time delay. This is surprisingly as one would expect such behaviour especially from the Dicke system, which is chaotic in nature \cite{Clive-Brandes_Chaos_and_qpt_Dicke}. However,the Dicke model has an additional bosonic mode which is not bounded by a conservation law like the spin components. This could be a reason, why no chaotic behaviour appears there. 

We think, that the feedback-induced limit cycles are hidden property of the LMG modell. On the one hand limit cycles are a natural property of  the  closed system. On the other hand, the time dependent limit cycles describe the ESQPT signal pretty well. Though the chaotic behaviour is an artefact of the time delay feedback, as it was neither a part of a closed LMG system. 

The feedback scheme is applied at a semiclassical level, thus we have neglected the influence of fluctuations in the thermodynamic limit, which could be important \cite{Dicke_open-critical_exponent_of_noise-Nagy}. Nevertheless, we think, that the fluctuations have not dramatic contributions to the described effects for $N \gg 1$ as they should scale with $1/\sqrt{N}$ and the used feedback scheme does not modify $\gamma$ in a strong way. Furthermore, the semiclassical LMG model predicts in many cases the same results \cite{ESQPT_in_quantum_optical_models-Pedro,brandes2013excited,LMG-TC-periodic_dynamic_and_QPT-Georg}. But the full quantum version of the considered feedback type  still remains an open issue. However, a recently published article \cite{Grimsmo-time_delayed_quantum_feedback_control} shows a way to go beyond mean field for a coherent type of feedback where the author describes feedback action via mapping to a bigger system. This would be one possible way to study the role of oscillations in quantum systems with one special feedback type. 

\textit{Acknowledgments. ---}
We thank Georg Engelhardt and Mathias Hayn for useful discussions. The authors gratefully acknowledge financial support from the DAAD and DFG Grants BR $1528/7-1$, $1528/8-2$, $1528/9-1$, SFB $910$, and GRK $1558$.

\appendix

\section{Boundary condition}
\label{ap:boundaries}
To determine the boundaries in Fig. \ref{fig1} we choose only imaginary $\Lambda$ values in Eq. \eqref{eq:stability_eq}, thus $\Lambda \equiv i \cdot s, s \in \mathbb{R}$. Splitting then the Eq. \eqref{eq:stability_eq} in an imaginary and real part, we obtain the correspondingly conditions

\begin{align}\label{eq:appendix_im-re-part}
0&=G_0 + G_1 \cdot \cos(s \tau) + G_2 \cdot \sin(s \tau), \\
0&=G_3 + G_4 \cdot \cos(s \tau) + G_5 \cdot \sin(s \tau), \notag \\
G_0 &= -\frac{\kappa  {(J_x^0)}^2 s}{{J_z^0}}+\frac{2 \gamma {J_x^0} J_y^0 s}{{J_z^0}}-4 {J_x^0} J_y^0 {J_z^0} \lambda  s \notag \\
	& -\frac{\kappa  {(J_y^0)}^2 s}{{J_z^0}}+2 \kappa  {J_z^0} s, \notag\\
G_1 &= 4 J_x^0 J_y^0 J_z^0 \lambda  s, \notag \\
G_2 &= -4 h {(J_x^0)}^2 J_z^0 \lambda -4 \kappa  J_x^0 J_y^0 {(J_z^0)}^2 \lambda, \notag \\
\end{align}
\begin{align}
G_3 &= h^2-s^2+\frac{2 \gamma h {(J_x^0)}^2}{J_z^0}-2 \gamma h J_z^0-\kappa ^2 {(J_x^0)}^2-\kappa ^2 		{(J_y^0)}^2 		\notag \\
	& +\kappa ^2 {(J_z^0)}^2-4 h {(J_x^0)}^2 J_z^0 \lambda -4 \kappa  J_x^0 J_y^0 {(J_z^0)}^2 \lambda, \notag \\
G_4&= 4 h {(J_x^0)}^2 J_z^0 \lambda +4 \kappa  J_x^0 J_y^0 {(J_z^0)}^2 \lambda, \notag \\
G_5 &= 4 J_x^0 J_y^0 J_z^0 \lambda  s. 
\end{align}

Bringing all $\sin$ and $\cos$  terms in both equations to one side, squaring them and adding together, we eliminate the $\tau$-dependence and obtain the following equation 

\begin{align}
0 &= s^4 + F_1 s^2 + F_0, \\
\intertext{with}
F_0&= \frac{1}{{(J_z^0)}^2}\biggl{\{}\bigg{(}h^2 J_z^0+2 \gamma h \left({(J_x^0)}^2-{(J_z^0)}^2\right) \notag \\
	& \quad\quad\quad +\kappa ^2 J_z^0 \left(-{(J_x^0)}^2-{(J_y^0)}^2+{(J_z^0)}^2\right) \bigg{)} \cdot \notag \\
	&\quad\quad\quad \bigg{(}h^2 J_z^0+2 h \left({(J_x^0)}^2 \left(\gamma-4 {(J_z^0)}^2 \lambda \right)-\gamma {(J_z^0)}^2\right) \notag \\
	&\quad\quad\quad  +\kappa  J_z^0 \left(-\kappa  {(J_x^0)}^2-8 J_x^0 J_y^0 {(J_z^0)}^2 \lambda +\kappa  \left({(J_z^0)}^2-{(J_y^0)}^2\right)\right) \notag \\
	& \quad \quad \quad \bigg{)}\biggl{\}},\notag
\end{align}
\begin{align}
F_1 &= \frac{1}{{(J_z^0)}^2}\biggl{\{}-2 h^2 {(J_z^0)}^2+4 h \left({(J_x^0)}^2 \left(2 {(J_z^0)}^3 \lambda -\gamma J_z^0\right)+\gamma {(J_z^0)}^3\right) \notag \\
    & +\kappa ^2 {(J_x^0)}^4-4 \kappa  {(J_x^0)}^3 J_y^0 \left(\gamma-2 {(J_z^0)}^2 \lambda \right) \notag \\
    & +2 {(J_x^0)}^2 \left({(J_y^0)}^2 \left(\kappa ^2+2 \gamma \left(\gamma-4 {(J_z^0)}^2 \lambda \right)\right)-\kappa ^2 {(J_z^0)}^2\right)\notag \\
    &-4 \kappa  J_x^0 J_y^0 \left({(J_y^0)}^2 \left(\gamma-2 {(J_z^0)}^2 \lambda \right)+2 {J_z^0}^4 \lambda -2 \gamma {(J_z^0)}^2\right) \notag \\
    &+\kappa ^2 \left({(J_y^0)}^4-2 {(J_y^0)}^2 {(J_z^0)}^2+2 {(J_z^0)}^4\right)\biggl{\}}, \notag 
\end{align}
which can be solved for $s$ and has then in general 4 different solutions

\begin{equation}
\label{eq:appendix_s-eq}
s = \pm \frac{1}{\sqrt{2}} \sqrt{-F_1 \pm \sqrt{F_1^2-4\cdot F_0}}. 
\end{equation}

The Eq. \eqref{eq:appendix_s-eq} fixes now the eigenvalue $\Lambda = i s$ for a given fixed point and feedback strength $\lambda$. Eq. \eqref{eq:appendix_im-re-part} gives for every fixed $s$-value the corresponding time-delay $\tau$, solving for example the imaginary part for $\tau$ we obtain
\begin{equation}
\tau = \frac{1}{s}\arctan\left(\frac{\frac{\pm\frac{G_1 \sqrt{-G_2^2 \left(G_0^2-G_1^2-G_2^2\right)}}{G_1^2+G_2^2}+\frac{G_0 G_1^2}{G_1^2+G_2^2}-G_0}{G_2}}
							{\frac{\pm\sqrt{-G_0^2 G_2^2+G_1^2 G_2^2+G_2^4}-G_0 G_1}{G_1^2+G_2^2}}
							\right) + \frac{2\pi}{s} \cdot z,
\end{equation}
where $z \in \mathbb{Z}$. The choice of $z$ is necessary to get boundary conditions at higher time delays $\tau$. Note, that the s dependence is also hidden in $G_i$.   

At this step we have still to much solutions. A lot of them are non-physical (if $\tau <0$ or $\tau \in \mathbb{C}$) or do not fulfil the real part equation Eq. \eqref{eq:appendix_im-re-part} and have to be sorted out. The remaining solutions are plotted as a brown line in Fig. \ref{fig1}.

\end{document}